# Coexistence of Terrestrial and Aerial Users in Cellular Networks

Mohammad Mahdi Azari[1], Fernando Rosas[2,3], Alessandro Chiumento[1], Sofie Pollin[1]

[1] Department of Electrical Engineering, KU Leuven, Belgium
[2] Centre of Complexity Science and Department of Mathematics, Imperial College London, UK
[3] Department of Electrical and Electronic Engineering, Imperial College London, UK

Email: mahdi.azari@kuleuven.be

*Abstract*—Enabling the integration of aerial mobile users into existing cellular networks would make possible a number of promising applications. However, current cellular networks have not been designed to serve aerial users, and hence an exploration of design parameters is required in order to allow network providers to modify their current infrastructure. As a first step in this direction, this paper provides an in-depth analysis of the coverage probability of the downlink of a cellular network that serves both aerial and ground users. We present an exact mathematical characterization of the coverage probability, which includes the effect of base stations (BSs) height, antenna pattern and drone altitude for various types of urban environments. Interestingly, our results show that the favorable propagation conditions that aerial users enjoy due to their altitude is also their strongest limiting factor, as it leaves them vulnerable to interference. This negative effect can be substantially reduced by optimizing the flying altitude, the base station height and antenna down-tilt angle. Moreover, lowering the base station height and increasing down-tilt angle are in general beneficial for both terrestrial and aerial users, pointing out a possible path to enable their coexistence.

## I. INTRODUCTION

Drone or unmanned aerial vehicle (UAV) based applications have been the subject of great interest in recent times. In effect, the community is starting to consider using UAVs for diverse scenarios such as search and rescue missions, data collection in Internet-of-Things (IoT), and remote location sensing. However, in order for this to become a reality, a fast and reliable connection between the UAV and a controller or data sink is a critical requirement. Therefore, the wireless aspect of UAV communication is gaining increasing interest in the research community, especially when the UAV is to be interfaced to an existing network for ubiquitous long-range connectivity [1].

A cost-effective way of satisfying the requirements for long-range reliability is by using an already existing and accessible technology, such as the ground cellular network. However, it is to be noted that the current cellular network was designed to serve users placed at the ground level or within buildings. As a matter of fact ground-to-drone communication is significantly different from traditional ground-to-ground links, as there is a strong dependency between the channel characteristics and the flying altitude [2]–[8]. As a consequence of this, a drone experiences more favorable propagation conditions as the altitude increases. Some consequences of this has been considered in [1], [9], where the feasibility of using the existing Long Term Evolution (LTE) infrastructure for UAV as an aerial user were studied. Results show that a UAV is able to receive signals from an increasing number of base stations as its height increases. Although these measurement-based works provide an interesting baseline on the network performance for drone operation at higher altitudes, it is not straightforward how to generalize their results in order to explore the impact of various fundamental system parameters.

Recent theoretical works on the propagation behavior for UAV communications have shown that the dependency between altitude and link quality can be modeled by combining the path loss and fading effects corresponding to LoS and non-LoS (NLoS) links, whose parameters might vary as function of the UAV altitude [5]–[8], [10], [11]. These models have been used to reflect the effects of altitude over the achievable performance of a wireless communication service. In [5], [7], the authors determine the optimal altitude for an aerial base station (ABS) as a result of a coverage area optimization process. In particular, [6] optimizes the altitude for maximum sum-rate and power gains, while [8] maximizes the coverage probability for a ground user by optimizing the drones flying altitude, density and antenna beamwidth.

The study of the propagation of wireless signals in traditional ground networks has a long history, generating a number of stochastic models which have been developed with the aid of extensive measurement campaigns [12]–[15]. Although such efforts are still to be made for the case of air-to-ground networks, from the existent literature it is clear that a base station to UAV link has much higher probability of LoS. Intuitively, this has the double effect of providing a stronger link to the serving base station while at the same time increasing the received interference. However, it is not clear which of these two effects is dominant. Furthermore, the design of current cellular networks did not consider mobile aerial nodes and hence base stations antennas are tilted downwards to maximize coverage at ground level. At this stage one might wonder if there exist a trade-off between providing coverage to ground or aerial users, which could be explored by varying the antenna down-tilt angle.

Our goal is to clarify the effect of including drones into existing cellular networks and explore if a satisfactory coexistence between ground and aerial nodes is possible. To achieve this aim, this paper combines the propagation models for aerial links with the traditional stochastic propagation tools used for cellular planning. In particular, we introduce a generic analytical framework to model the coverage probability of a UAV user equipment from a ground network base station. An LoS/NLoS propagation model is used for both path loss and small scale fading, and the effects for different types of urban environments (i.e. Suburban, Urban, Dense Urban, and Highrise Urban) are analyzed by including a generic distance and height-dependent LoS probability [16]. After deriving an exact expression for the coverage probability, the proposed framework allows to study how the different urban propagation environments and various network parameters, such as the UAV altitude, the ground base station height and its antenna down-tilt impact the coverage performance. Our results show that the coexistence of aerial and ground UEs is highly non-trivial, as for different environments the variation of specific network parameters act in opposite ways on the performance of aerial and ground users. However, a larger antenna down-tilt seems to be beneficial for both ground and aerial UEs.

The remainder of this paper is structured as follows. Section II presents the system model, it includes both the cellular network architecture considered and the channel models. In Section III the coverage probability formulation is obtained and an exact expression is derived and in Section IV the numerical simulation results are presented and discussed. Finally, in Section V, the final conclusions are drawn.

## II. NETWORK MODEL

First, the considered cellular network architecture is introduced in Section II-A. Section II-B describes the channel model and, finally, in Section II-C the UE to BS association method selected in this work is discussed and the relative signal-to-interference-ratio (SIR) of each link is quantified.

### A. Cellular Network Architecture

We consider a cellular downlink network consisting of multiple ground base stations (BSs) placed at the same height $h_{BS}$, and a drone as an aerial UE located at altitude $h_D$. The ground BSs are randomly distributed according to a homogeneous Poisson point process (HPPP) $\Phi$ of a fixed density $\lambda$ BSs/km$^2$. For simplicity, instead of sectored base stations, ominidirectional horizontal antenna patterns are considered. The BS vertical antenna pattern is, on the other hand, directional to account for the down-tilt. The vertical antenna beamwidth and down-tilt angle of the BSs are respectively denoted by $\theta_B$ and $\theta_t$, as illustrated in Figure 1. The side and main lobe gains of the antennas are denoted by $G_s$ and $G_m$ respectively.

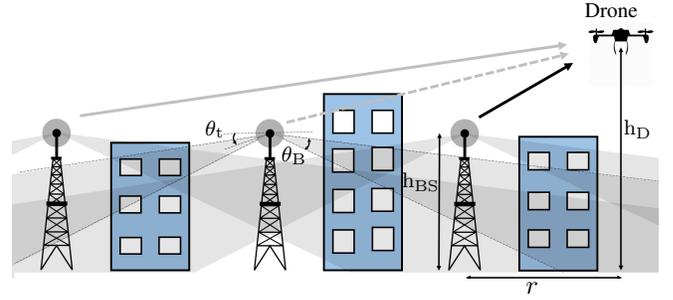

Fig. 1. Ground cellular network and user equipment at altitude $h_D$.

The aerial and ground UEs, on the other hand have omnidirectional antennas. The distance between the projection of the UAV on the ground and the $i$-th BS is represented by $r_i$. This results in a communication link length $d_i = \sqrt{r_i^2 + (h_D - h_{BS})^2}$.

### B. Channel Model

In order to model the wireless channel between the $i$th ground BS and a UE, the LoS and non-line-of-sight (NLoS) components are considered separately along with their probabilities of occurrence. To this end, the path losses for each component can be expressed as

$$\zeta_v(r_i) = A_v d_i^{-\alpha_v} = A_v \left[ r_i^2 + (h_{BS} - h_D)^2 \right]^{-\alpha_v/2}, \quad (1)$$

where $v \in \{L, N\}$, $\alpha_L$ and $\alpha_N$ are the path loss exponents for the LoS and NLoS links respectively, and $A_L$ and $A_N$ are constants representing the path losses at the reference distance $d_i = 1$ for the LoS and NLoS cases respectively.

Moreover, each channel suffers from a small scale fading with $\Omega_L$ and $\Omega_N$ being the fading powers for the LoS and NLoS links respectively. A Nakagami-m fading model is selected which can represent various fading environments. Accordingly, the channel gains $\Omega_L$ and $\Omega_N$ follow a gamma distribution with the probability density function (PDF) expressed as [17]

$$f_{\Omega_v}(\omega) = \frac{m_v^{m_v} \omega^{m_v - 1}}{\Gamma(m_v)} \exp(-m_v \omega); \quad v \in \{L, N\}, \quad (2)$$

where $m_L$ and $m_N$ are the fading parameters for the LoS and NLoS links respectively, assumed to be integers for analytical tractability.

Therefore, by considering that all the BSs transmit at the same power level $P_t$, the received power $P_r$ at the drone from the LoS and NLoS components can be expressed as

$$P_r(r_i) = \begin{cases} P_t\, G(r_i)\, \zeta_L(r_i)\, \Omega_L & ; \text{ for LoS} \\ P_t\, G(r_i)\, \zeta_N(r_i)\, \Omega_N & ; \text{ for NLoS} \end{cases} \quad (3)$$

where $G(r_i)$ represents the effect of transmitter antenna's gain.

Finally, the probability of LoS $\mathcal{P}_\text{L}$ between a BS at a ground distance $r_i$ from the drone in an urban environment is given by [16]

$$\mathcal{P}_\text{L}(r_i) = \prod_{n=0}^{m}\left[1-\exp\left(-\frac{\left[\text{h}_\text{BS}-\frac{(n+0.5)(\text{h}_\text{BS}-\text{h}_\text{D})}{m+1}\right]^2}{2c^2}\right)\right],\tag{4}$$

where $m = \lfloor\frac{r_i\sqrt{ab}}{1000}-1\rfloor$. In this model, an urban area is defined as a set of buildings placed in a square grid in which $a$ is the fraction of the total land area occupied by the buildings, $b$ is the mean number of buildings per km$^2$, and the buildings height is modeled by a Rayleigh PDF with an scale parameter $c$. Please note that the proposed expression for LoS probability in (4) is a decreasing step function of $r_i$ and an increasing function of $\text{h}_\text{D}$. Finally, the probability of NLoS is $\mathcal{P}_\text{N}(r_i) = 1 - \mathcal{P}_\text{L}(r_i)$.

### C. Base Station Association and Link SIR

In this work, we focus on ground cellular networking, the case where the drone associates with the closest BS. Accordingly, the communication link between the drone and the associated BS is interfered by all the other neighboring BSs. By considering the channel model described in Section II-B, the aggregate interference can be written as

$$I = \sum_{i\in\Phi\backslash\{0\}}\text{P}_\text{r}(r_i),\tag{5}$$

where index 0 belongs to the closest BS. Now, assuming that the noise power is negligible compared to the aggregate interference, the signal-to-interference-ratio (SIR) at the drone is

$$\text{SIR} = \frac{\text{P}_\text{r}(r_0)}{I} = \begin{cases}\frac{\text{P}_\text{t}\,G(r_0)\,\zeta_\text{L}(r_0)\,\Omega_\text{L}}{\sum_{i\in\Phi\backslash\{0\}}\text{P}_\text{r}(r_i)} & ;\text{ for LoS} \\ \frac{\text{P}_\text{t}\,G(r_0)\,\zeta_\text{N}(r_0)\,\Omega_\text{N}}{\sum_{i\in\Phi\backslash\{0\}}\text{P}_\text{r}(r_i)} & ;\text{ for NLoS}\end{cases}\tag{6}$$

where $r_0$ is a random variable representing the ground distance between the drone and the closest BS.

## III. COVERAGE PROBABILITY

In this section the coverage probability for ground and aerial users are derived; Then, the result will be instantiated for a Rayleigh fading assumption which enables an exact closed-form expression. This assumption is mainly justified for NLoS communication link and is widely adopted due to tractability.

The coverage probability of the link between a drone-UE and its associated BS can be defined as follows

$$\mathcal{P}_\text{cov} = \mathbb{P}[\text{SIR} > \text{T}],\tag{7}$$

where T is an SIR threshold and SIR is expressed in (6). By considering the dependency of the location of serving BS at $r_0$ on the underlying HPPP, we can express (7) as

$$\mathcal{P}_\text{cov} = \int_0^\infty \mathbb{P}[\text{SIR}>\text{T}|r_0]\,f_{\text{R}_0}(r_0)\,dr_0,\tag{8a}$$

$$= \int_0^\infty \left[\mathcal{P}^\text{L}_{\text{cov}|r_0}\cdot\mathcal{P}_\text{L}(r_0) + \mathcal{P}^\text{N}_{\text{cov}|r_0}\cdot\mathcal{P}_\text{N}(r_0)\right]\\\times f_{\text{R}_0}(r_0)\,dr_0,\tag{8b}$$

where

$$\mathcal{P}^\text{L}_{\text{cov}|r_0} = \mathbb{P}[\text{SIR}>\text{T}|r_0,\text{LoS}],\tag{9a}$$
$$\mathcal{P}^\text{N}_{\text{cov}|r_0} = \mathbb{P}[\text{SIR}>\text{T}|r_0,\text{NLoS}],\tag{9b}$$

and $f_{\text{R}_0}(r_0)$ is the PDF of the closest BS's ground distance from the UAV which, according to standard results for the HPPP [12] can be expressed as

$$f_{\text{R}_0}(r_0) = 2\pi\lambda r_0\exp(-\lambda\pi r_0^2).\tag{10}$$

In (9) $\mathcal{P}^\text{L}_{\text{cov}|r_0}$ and $\mathcal{P}^\text{N}_{\text{cov}|r_0}$ are the conditional coverage probabilities for the LoS and NLoS of the desired link (between the drone and the closest BS) respectively given the distance $r_0$. In the following lemma we derive expressions for the conditional coverage probabilities.

**Lemma 1.** *The conditional coverage probabilities of the LoS and NLoS links can be obtained as*

$$\mathcal{P}^v_{\text{cov}|r_0} = \sum_{k=0}^{m_v-1}\frac{(-s_v)^k}{k!}\frac{d^k}{ds_v^k}\mathcal{L}_{I|r_0}(s_v),\quad v\in\{\text{L},\text{N}\}\tag{11a}$$

*where*

$$\mathcal{L}_{I|r_0}(s_v) = \exp\left(-2\pi\lambda\int_{r_0}^\infty[1-\Upsilon_\text{L}(r,s_v)\cdot\mathcal{P}_\text{L}(r)\\-\Upsilon_\text{N}(r,s_v)\cdot\mathcal{P}_\text{N}(r)]\,r\,dr\right),\tag{11b}$$

*and*

$$s_v = \frac{m_v\text{T}}{\text{P}_\text{t}\,G(r_0)\,\zeta_v(r_0)},\tag{11c}$$

$$\Upsilon_\text{L}(r,s_v) = \left(\frac{m_\text{L}}{m_\text{L}+s_v\text{P}_\text{t}\,G(r)\,\zeta_\text{L}(r)}\right)^{m_\text{L}},\tag{11d}$$

$$\Upsilon_\text{N}(r,s_v) = \left(\frac{m_\text{N}}{m_\text{N}+s_v\text{P}_\text{t}\,G(r)\,\zeta_\text{N}(r)}\right)^{m_\text{N}}.\tag{11e}$$

*Proof.* Please find Appendix A. □

By using (8)–(11) the total coverage probability is obtained as a function of $\text{h}_\text{D}$, $\text{h}_\text{BS}$, $\lambda$, $\theta_\text{B}$, $\theta_\text{t}$ and the type of environment, and given in the following theorem.

**Theorem 1.** *The coverage probability $\mathcal{P}_\text{cov}$ of the communication link between a drone-UE and the closest BS is obtained as follows*

$$\mathcal{P}_\text{cov} = 2\pi\lambda\int_0^\infty r_0\Big[\mathcal{P}_\text{L}(r_0)\sum_{k=0}^{m_\text{L}-1}\frac{(-s_\text{L})^k}{k!}\frac{d^k}{ds_\text{L}^k}\mathcal{L}_{I|r_0}(s_\text{L})\\+\mathcal{P}_\text{N}(r_0)\sum_{k=0}^{m_\text{N}-1}\frac{(-s_\text{N})^k}{k!}\frac{d^k}{ds_\text{N}^k}\mathcal{L}_{I|r_0}(s_\text{N})\Big]e^{-\lambda\pi r_0^2}\,dr_0.\tag{12}$$

For the particular case of $m_\text{L} = m_\text{N} = 1$, which has been of interest in the literature, the above formula finds a simple expression that is shown in the next corollary.

**Corollary 1.** *Assuming Rayleigh fading for both LoS and NLoS links, i.e. $m_L = m_N = 1$, the coverage probability is given by*

$$\mathcal{P}_{\text{cov}} = 2\pi\lambda \int_0^\infty r_0 \exp(-\lambda\pi r_0^2)[\mathcal{P}_L(r_0)\mathcal{L}_{I|r_0}(s_L) + \mathcal{P}_N(r_0)\mathcal{L}_{I|r_0}(s_N)] \, dr_0. \quad (13)$$

## IV. RESULTS AND DISCUSSIONS

In this section, we use our proposed framework to analyze the effects of different network design parameters on the coverage probability of both aerial and terrestrial UEs. The results allow us to make recommendations to improve the connectivity of drones deployed in current and future cellular networks. The simulation parameters and their default values are listed in Table I. They are chosen in such a way to reflect as much as possible realistic cellular deployment parameters.

TABLE I. Numerical result and simulation parameters

| Parameter | Value |
|---|---|
| $(\alpha_L, \alpha_N)$ | (2.09, 3.75) |
| $(A_L, A_N)$ | $(-41.1, -32.9)$ dB |
| $(m_L, m_N)$ | (1, 3) |
| $P_t$ | $-6$ dB |
| T | 0.3 |
| $(a, b, c)$ | (0.3, 500, 15) |
| $\lambda$ | 50 |
| $(\theta_B, \theta_t)$ | $(40^o, 30^o)$ |
| $(G_m, G_s)$ | (10, 0.5) |
| $h_D$ | 60 m |

### A. Impact of Small Scale Fading and Drone Altitude

In general, the coverage probability decreases with the drone altitude. This is because of a rise in interference due to the increasingly dominant effect of LoS links between the UAV and the ground base stations. There is a slight increase in coverage probability for very modest altitudes, due to the LoS link between the UAV and the closest (and serving) base station while, on the other hand, the UAV has not risen high enough to decrease the NLoS probability with the further base stations.

It is also visible from Figure 2 that the small scale fading influences consistently the coverage probability. Results are obtained by simulating different fading models: a pure Rayleigh propagation ($m_L = m_N = 1$), a propagation in which there is no fading ($m_N, m_L \to \infty$) obtained by imposing $m_L = m_N = 100$ which is large enough [18] and, finally a propagation model with $m_L = 3, m_N = 1$. The Rayleigh assumption underestimates the actual coverage probability up to around $h_D = 80$m, however higher than that the Rayleigh assumption overestimates $\mathcal{P}_{\text{cov}}$.

### B. Impact of Base Station Height and Different Environments

Interestingly, our results suggest that increasing the BS height generally deteriorates the coverage performance for both ground-UE and drone-UE. In fact, as the BS height is increased, the LoS component becomes prominent on both the interference and the signal paths. The growth in the interference power, however, is dominant and hence leads to a reduction in coverage. This trend, especially for a ground-UE as shown in Figure 3a, depends on the type of environment and the BS antenna pattern. In particular, for a ground-UE in a propagation environment with many obstacles (such as a dense urban or high-rise urban scenarios), an increase in BS height is generally beneficial. In high-rise urban environment, the increase in the signal power with the BS height is dominant over the interference up to 20m, while if the base station is higher than that the transition of the interference paths from NLoS to LoS causes the coverage performance to decrease. However, for a drone-UE in a very densely built up environment coverage performance is robust to BS height since the LoS probability does not significantly vary for the examined range.

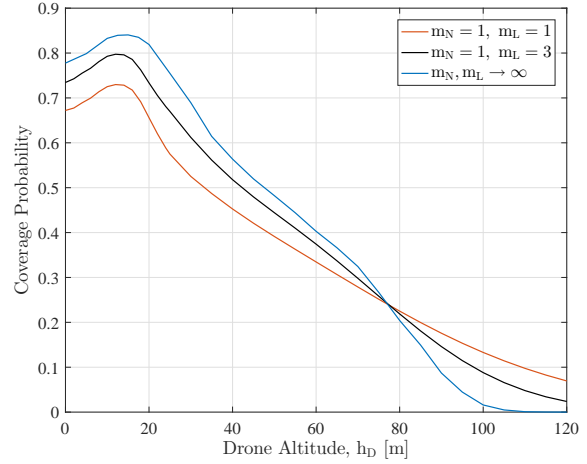

Fig. 2. Coverage probability with respect to drone's altitude and different fading parameters.

In contrast to a ground-UE, for a drone-UE increasing the BS height does not affect the coverage after an environment-dependent $h_{BS}$ value. This is due to the fact that both the serving and dominant interfering BSs become LoS. This effect is further emphasized in a High-rise Urban environment, in which an increase in BS height in the beginning does not significantly change the LoS condition to the drone resulting in a constant coverage probability up to a certain height. Note that a more densely built environment leads to a higher performance due to LoS blocking. In other words, in the presence of more blockages the interference level is reduced, improving the coverage performance in turn.

The impact of BSs height on the coverage and the above-mentioned trend is also dependent on the BS antenna down-tilt $\theta_t$. For instance, Figure 3a illustrates that for the case of high-rise urban and $\theta_t = 15^o$, the coverage probability of a ground user continuously decreases in contrast to the case in which $\theta_t = 30^o$. Our results suggest that a higher down-tilt angle (i.e. high $\theta_t$), in which the BS antenna's main lobe is lowered towards the ground, provides better coverage.

The effect of the down-tilt is particularly strong over drone-UEs, as illustrated by Figure 3b. When a UAV is at altitudes above $h_{BS}$, drone-UEs are reached by the main lobes of the interfering BSs while still experiencing the side lobes of its serving base station, and hence reducing the coverage probability.

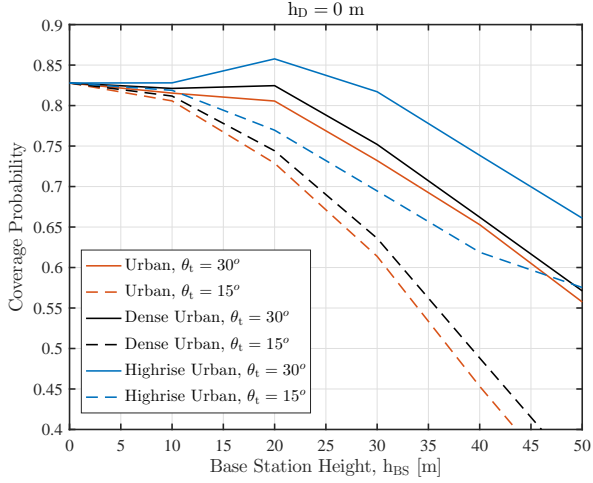

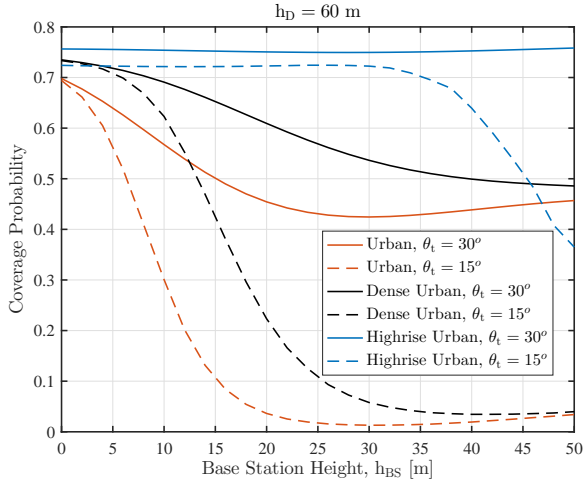

Fig. 3. Coverage Probability as a function of base station's height and down-tilt angle for (a) ground user and (b) aerial user.

Furthermore, while the coverage probability for a ground user is monotonically increasing with the down-tilt, the drone-UE's performance as function of the down-tilt, is also extremely dependent on its altitude when compared with the base station's height. The effect of BS antenna down-tilt on an aerial UE is highly non linear with the drone's altitude as both parameters dictate when the interfering base stations' main lobes reach the UAV, increasing thus the LoS condition to the interferers. This shows that, for any down-tilt angle, it is beneficial for a drone-UE to fly close to BS height as visible in Figure 4.

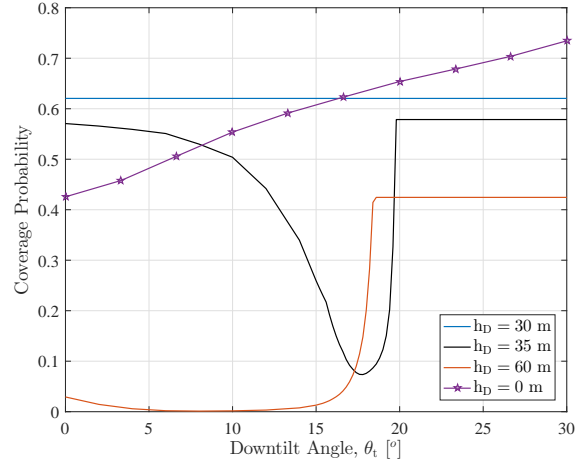

Fig. 4. BS antenna down-tilt effects on the coverage probability with respect to user altitude.

## V. CONCLUSION

We have analyzed the key issues related to the integration of low altitude drone-UEs into existent wireless cellular networks. In particular, we presented a generic framework for evaluating the coverage performance in a network that includes air-to-ground and ground-to-ground communication links. This framework considers the impact of fundamental design parameters such as BS height, antenna pattern and drone altitude for different types of environments. Within this framework, we presented the derivation of an exact expression for the coverage probability, which can be used to develop some first insights about the impact of the above mentioned parameters, clarifying fundamental issues and trade-offs of the coexistence of aerial and terrestrial users.

Our results show that the beneficial propagation conditions an aerial user experiences at high altitudes actually has an overall negative effect, as the high vulnerability to interference dominates over the increased received signal power. Therefore, restricting the flying altitudes could be beneficial for the communication capabilities. We also found that lowering the BS height and increasing the BS down-tilt angle might improve the performance of both ground and drone UEs. We hope that this unexpected win-win condition might serve as a first step in enabling a satisfactory integration of these two technologies in the future.

## APPENDIX

Let us calculate $\mathcal{P}^v_{\text{cov}|r_0}$ using the definition (9) and the expression for SIR given in (6). One can start by noting that

$$\mathcal{P}^v_{\text{cov}|r_0} = \mathbb{P}\left[\frac{P_t\, G(r_0)\, \zeta_v(r_0)\, \Omega_v}{I} > T\right]$$

$$= \mathbb{E}_I\left\{\mathbb{P}\left[\Omega_v > \frac{T}{P_t\, G(r_0)\, \zeta_v(r_0)}I\right]\right\}$$

$$\stackrel{(a)}{=} \mathbb{E}_I\left\{\sum_{k=0}^{m_v-1} \frac{s_v^k}{k!} I^k \exp(-s_v I)\right\}, \quad (14)$$

where (a) follows from the gamma distribution of $\Omega_v$ with an integer parameter $m_v$, and $s_v$ is expressed in (11c). Therefore, we can write

$$\mathcal{P}_{\text{cov}|r_0}^v = \sum_{k=0}^{m_v-1} \frac{s_v^k}{k!} \cdot \mathbb{E}_I \left\{ I^k \exp(-s_v I) \right\}$$

$$= \sum_{k=0}^{m_v-1} \frac{(-s_v)^k}{k!} \cdot \frac{d^k}{ds_v^k} \mathcal{L}_{I|r_0}(s_v), \quad (15)$$

where

$$\mathcal{L}_{I|r_0}(s_v) = \mathbb{E}_I\{\exp(-s_v I)\}$$

$$\stackrel{(a)}{=} \mathbb{E}_{\Phi,\Omega} \left\{ \prod_{i \in \Phi \setminus \{0\}} \exp(-s_L P_r(r_i)) \right\}$$

$$= \mathbb{E}_\Phi \left\{ \prod_{i \in \Phi \setminus \{0\}} \mathbb{E}_\Omega \left\{ \exp(-s_v P_r(r_i)) \right\} \right\}$$

$$\stackrel{(b)}{=} \mathbb{E}_\Phi \left\{ \prod_{i \in \Phi \setminus \{0\}} \mathbb{E}_\Omega \{\exp(-s_v P_r^L(r_i)) \cdot \mathcal{P}_L(r_i) \right.$$

$$\left. + \exp(-s_v P_r^N(r_i)) \cdot \mathcal{P}_N(r_i) \} \right\}$$

$$\stackrel{(c)}{=} \mathbb{E}_\Phi \left\{ \prod_{i \in \Phi \setminus \{0\}} [\Upsilon_L(r_i, s_v) \cdot \mathcal{P}_L(r_i) \right.$$

$$\left. + \Upsilon_N(r_i, s_v) \cdot \mathcal{P}_N(r_i)] \right\}$$

$$\stackrel{(d)}{=} \exp\left( -2\pi\lambda \int_{r_0}^{\infty} [1 - \Upsilon_L(r, s_v) \cdot \mathcal{P}_L(r) \right.$$

$$\left. - \Upsilon_N(r, s_v) \cdot \mathcal{P}_N(r)] \, r \, dr \right). \quad (16)$$

Above, (a) follows from (5), in (b)

$$P_r^L(r_i) = P_t\, G(r_i)\, \zeta_L(r_i)\, \Omega_L,$$
$$P_r^N(r_i) = P_t\, G(r_i)\, \zeta_N(r_i)\, \Omega_N,$$

in (c), $\Upsilon_L$ and $\Upsilon_N$ defined in (11d) and (11e) respectively are used, and (d) is obtained using the probability generating functional (PGFL) of PPP. Please note that PGFL for a general point process $\Phi$ is defined as

$$\text{PGFL} = \mathbb{E}\left\{ \prod_{x \in \Phi} f(x) \right\}, \quad (17)$$

and in particular for a PPP of density $\lambda$ it is equal to

$$\text{PGFL} = \exp\left( -\lambda \int_A 1 - f(x)\, dx \right). \quad (18)$$